\documentclass[preprint,12pt]{elsarticle}
\usepackage{stywhispers,amsmath,epsfig}
\usepackage{amsmath,amsfonts,amssymb}
\usepackage{graphicx,makecell}
\usepackage[colorlinks=true, allcolors=blue]{hyperref}
\usepackage{subcaption}
\usepackage{dirtytalk}
\usepackage{tabularray}
\usepackage{float}
\usepackage{float}
\usepackage{amsmath}
\usepackage{booktabs}
\usepackage{xcolor}
\usepackage{colortbl}
\usepackage{makecell}
\usepackage{multirow}

\title{Hyperspectral Unmixing using Iterative, Sparse and Ensambling Approaches for Large Spectral Libraries Applied to Soils and Minerals}
\twoauthors
  {Jade Preston}
	{University of Virginia\\
	School of Data Science\\
	Charlottesville, VA}
  {William Basener}
	{University of Virginia\\
	School of Data Science\\
	Charlottesville, VA}
\begin{document}
%
\maketitle
\begin{abstract}
Unmixing is a fundamental process in hyperspectral image processing in which the materials present in a mixed pixel are determined based on the spectra of candidate materials and the pixel spectrum.  Practical and general utility requires a large spectral library with sample measurements covering the full variation in each candidate material as well as a sufficiently varied collection of potential materials.   However, any spectral library with more spectra than bands will lead to an ill-posed inversion problem when using classical least-squares regression-based unmixing methods. Moreover, for numerical and dimensionality reasons, libraries with over 10 or 20 spectra behave computationally as though they are ill-posed. In current practice, unmixing is often applied to imagery using manually-selected materials or image endmembers.  General unmixing of a spectrum from an unknown material with a large spectral library requires some form of sparse regression; regression where only a small number of coefficients are nonzero. This requires a trade-off between goodness-of-fit and model size.  In this study we compare variations of two sparse regression techniques, focusing on the relationship between structure and chemistry of materials and the accuracy of the various models for identifying the correct mixture of materials present. Specifically, we examine LASSO regression and ElasticNet in contrast with variations of iterative feature selection, Bayesian Model Averaging (BMA), and quadratic BMA (BMA-Q) --- incorporating LASSO regression and ElasticNet as their base model. To evaluate the the effectiveness of these methods, we consider the molecular composition similarities and differences of substances selected in the models compared to the ground truth.
\end{abstract}

\begin{keywords}
Hyperspectral Unmixing, Linear Regression, Regularization, Sparse Regression, Ensambling, Physical-chemical Phenomenon, Bayesian

\end{keywords}

\section{INTRODUCTION}
\label{Introduction}
Hyperspectral imaging is an imaging technology that provides analytical chemical information at a per-pixel level. This imagery consists of measurements of electromagnetic radiation, which is then converted to percent reflectance per-band of the materials in the image \cite{shaw2003spectral,campbell2011introduction,qian2021hyperspectral}. Hyperspectral imagers often collect spectral data mapped over hundreds of bands \cite{wei2020overview} providing information that is important in various fields of study from earth to space sciences as well as applications in medicine, military use and landscape development \cite{bioucas2013hyperspectral, borsoi2021spectral}. In particular, the spectrum for each pixel enables identification of the material or materials present in the pixel.

The process of determining the relative abundances of materials present in the area represented in a pixel spectrum is called spectral unmixing \cite{keshava2002spectral}.  If the materials are known \emph{a priori} (or the set of potential materials is known \emph{a priori} and small in number), then unmixing is a straightforward regression problem. If the set of potential materials is large, then unmixing includes identification of the materials present (from a large set of potential materials) as well as determining their abundances.  

In this paper, we address this substantially more complex identification-unmixing problem when the set of potential materials' spectra (called a spectral library) is large.  We show that ordinary least squares regression is guaranteed to fail, and then present and evaluate regression methods that have potential effectiveness for the identification-unmixing problem. We additionally investigate the effectiveness of the identification with respect to mineral chemistry and general mineral class.
\begin{figure}[ht] 
   \begin{center}
\includegraphics[width=.48\textwidth]{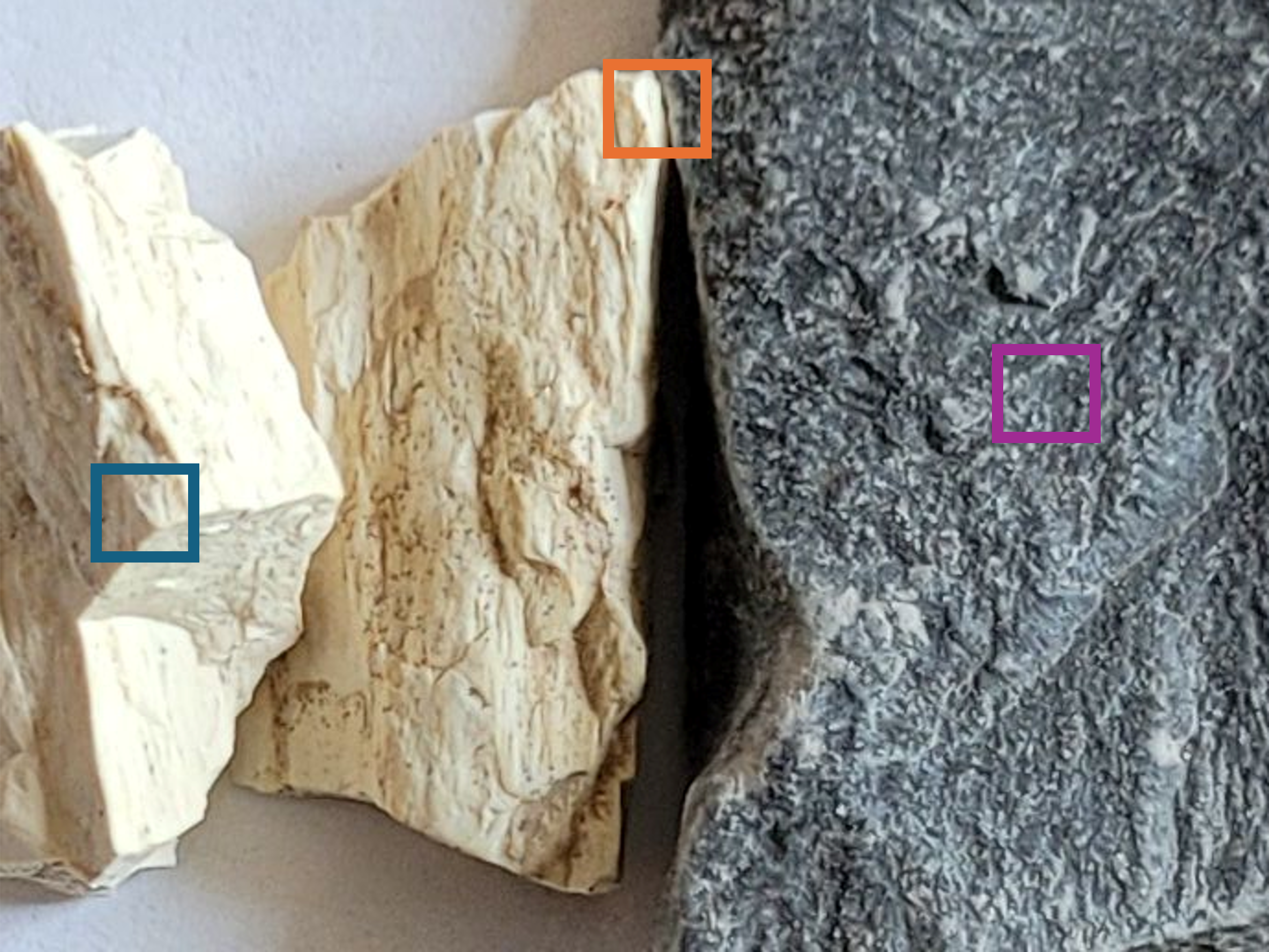}
	\end{center}
 \caption{Display of alunite (on the left) and kaolinite (on the right) minerals. The blue, purple and orange boxes represent pixels of alunite, kaolinite and a linear mixture of the two minerals respectively.}
 \label{fig:SimpleAluniteKaolinte}
\end{figure}

\subsection{Mixed Pixel Phenomenology}
In collecting hyperspectral imagery, there is a trade-off between spectral resolution, spatial resolution, and the signal-to-noise ratio (SNR). Increasing the spectral resolution (capturing finer spectral detail with narrower bands) or increasing the spatial resolution (capturing finer spatial detail with smaller pixels) results in lower SNR. Because the primary use of hyperspectral imagery is to provide useful spectral information (high SNR, high spectral resolution), these images have a relatively low spatial resolution \cite{borsoi2021spectral,keshava2002spectral}. A low spatial resolution means that the surface area corresponding to each pixel is large.  Spatial resolution is often quantified using the ground sample distance (GSD), which is the distance between the centers of the areas measured in neighboring pixels. Additionally, hyperspectral sensors are often designed to collect imagery covering large areas on the earth, which additionally creates the need for larger GSD. This large GSD means that each pixel is a measurement of a large area on the ground which often contains multiple materials that result in a mixed pixel \cite{shaw2003spectral,wei2020overview,ince2023spatial}. 

The mixture of materials in the area represented by a pixel can be categorized depending on whether the pixel spectrum will be a linear mixture or nonlinear mixture of the spectra for the individual constituents. Linear mixing occurs when the area measured in the pixel spectrum consists of more than one material and light reflects off only one of the materials before reaching the sensor.  This is common when the materials occupy adjacent regions in the area for the pixel. The abundance of each constituent spectrum in the pixel spectrum corresponds to the fractional abundance of area in the pixel occupied by the material~\cite{bioucas2012hyperspectral}. The orange box in the top-center of Figure~\ref{fig:SimpleAluniteKaolinte} provides an example of an area for a pixel that would result in a linear mixture of the minerals alunite and kaolinite. Although the linear mixing model is relatively simple, this model has fostered a huge amount of research leading to \say{a plethora of unmixing algorithms} \cite{bioucas2012hyperspectral}. 

Nonlinear mixing occurs when the area measured in the pixel spectrum consists of more than one material, and the physical arrangement of the materials results in a more complex interaction of light with the materials~\cite{bioucas2012hyperspectral, keshava2002spectral, keshava2003survey}.  This commonly occurs when the individual materials are powders or grains, and the pixel area involves particles from multiple materials mixed together. In this arrangement, called an intimate mixture \cite{keshava2002spectral}, light makes multiple bounces off of (and/or transmission through) different materials in the pixel area before reaching the sensor.  Another nonlinear mixture can also occur when light passes through vegetation leaves and reflects off materials underneath before returning to the sensor, creating a canopy effect \cite{keshava2002spectral}. While most unmixing models in practice focus on the more well-defined linear mixing case, there are numerous proposed methods for nonlinear unmixing \cite{heylen2014review}. 

\begin{figure*}[h!]
    \centering
    \includegraphics[width=.9\linewidth]{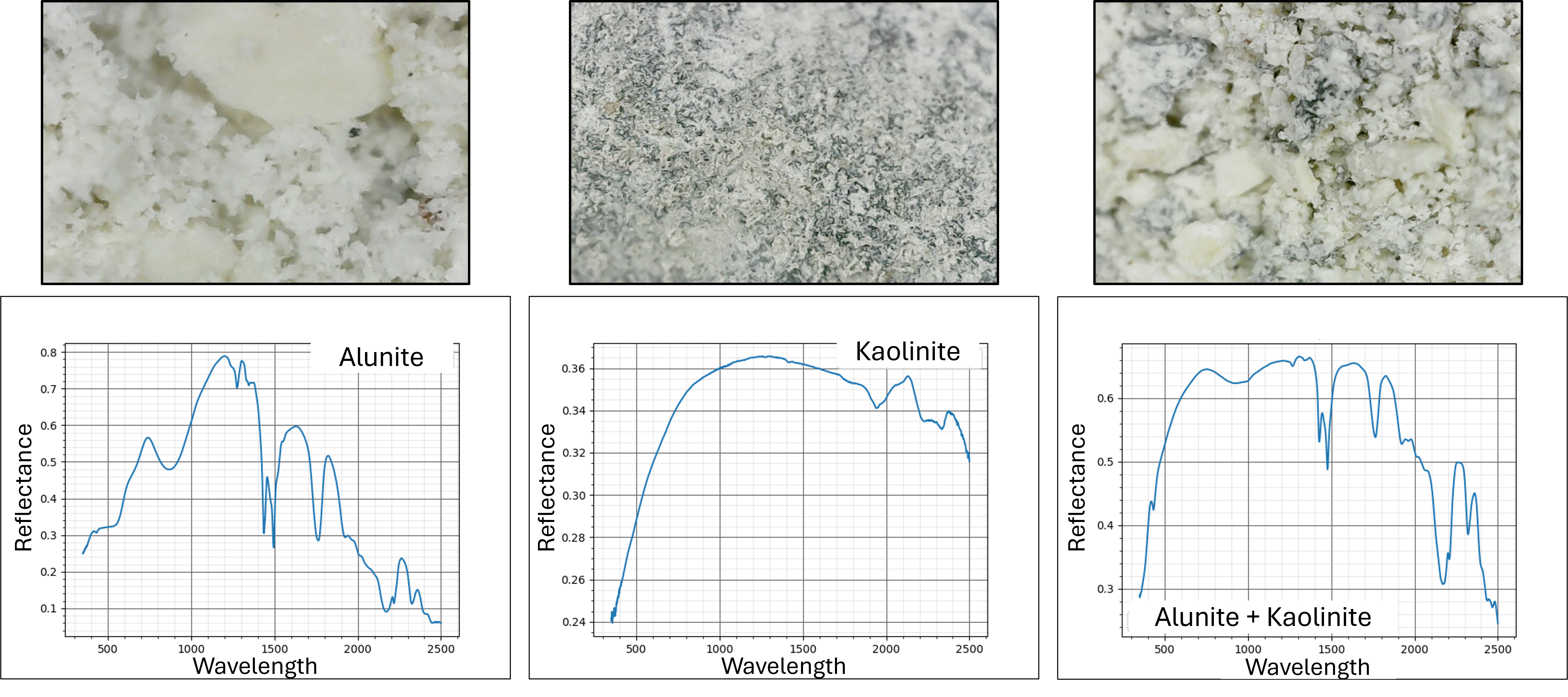}
    \caption{ Display of images (from left to right) of granular mineral compositions: alunite, kaolinite and a mixture of the two minerals. Underneath each image is their associated pixel spectrum.}
    \label{fig:complexmix}
\end{figure*}

An example of nonlinear mixing is shown in Figure~\ref{fig:complexmix}. To create this figure, the authors ground the alunite and kaolinite samples shown in Figure~\ref{fig:SimpleAluniteKaolinte} to obtain granular alunite (top-left) and kaolinite (top-center).  Spectra of these granular materials were measured with an ASD field spectrometer (2151 bands from 350-2500 nm) and are shown in Figure~\ref{fig:complexmix} (bottom-left and bottom-center, respectively).  The granular materials were mixed together vigorously. Their mixture and resulting nonlinear-mixed spectrum are shown in the right-hand side of Figure~\ref{fig:complexmix}.

\subsection{Hyperspectral Unmixing}
The mixed-pixel phenomenology is the process of creating a pixel spectrum that is a function of the spectra for the individual materials and their physical arrangement.  Unmixing is the inverse process, determining the constituent materials present in the mixture from the mixed pixel spectrum and the spectra of potential constituents \cite{borsoi2021spectral,keshava2000algorithm}.  The set of spectra for potential constituent materials is called a spectral library.  These spectra are often measurements from known materials.  

In general, unmixing requires choosing a statical model representing the function by which the constituent spectra are combined to create the resulting mixed spectrum and a loss function that is minimized to select the constituents from the library that are present and coefficients.  The loss function usually has terms for goodness-of-fit such as the sum of squares error along with terms for the number or quantity of constituents.  Sometimes a method for minimizing the loss function is also considered as part of the model.  The library spectra that the model determines to be present in the mixture are called inferred constituent spectra, and the output of the model that approximate the mixed pixel is called the inferred spectrum.

Unmixing is sometimes conducted by unmixing pixel spectra in an image using pixels from the same image as the potential constituents.  In this context, the potential constituent spectra from image pixels are usually called endmembers \cite{bioucas2012hyperspectral}.  Thus, this involves two steps: a small number of `pure' pixels are selected from an image (either manually or using an automated endmember selection algorithm) and then every pixel in the image is unmixed using these endmembers.  Unmixing an image using in-scene pixel endmembers in this way does not provide the level of information and certainty that unmixing with a library of known materials can provide.

Unmixing assuming a linear mixture model (linear unmixing) has lower variance than unmixing with a nonlinear model (nonlinear unmixing), and thus is a more stable process and is easier to interpret \cite{keshava2002spectral,manolakis2003hyperspectral,celik2023qlsu}. In this model, photons are assumed to be reflected off only one material at a time before returning to the sensor \cite{dobigeon2013nonlinear}. Figure \ref{fig:SimpleAluniteKaolinte} serves as a useful example where the linear mixture model can be readily employed. In this classic scenario, the observed mixed pixel spectrum -- from the orange box -- can be modeled by the linear mixture $\mathbf{y} = a_a \mathbf{s_a} + a_k \mathbf{s_k}$ in which $\mathbf{s_a}$ and $\mathbf{s_k}$ represent alunite and kaolinite spectra, and $a_a$ and $a_k$ depict the fractional abundances or proportion of space within the pixel that each mineral covers \cite{heinz2001fully}.

Least squares linear regression is the common interpretation of the linear mixture model and serves as the foundation for most unmixing models \cite{keshava2003survey}. The observed pixel spectrum can be represented by Equation \ref{eq:LMM_equation} where $\mathbf{y}$ is a vector of size \emph{W}x1 with \emph{W} bands, $\mathbf{S}$ is a matrix of size \emph{W}x\emph{N} with \emph{N} constituent spectra from the spectral library, $\mathbf{s_i}$ represents an individual spectra in $\mathbf{S}$, $\mathbf{a}$ is a vector of size \emph{N}x1 representing the coefficients or fractional abundances associated with each constituent spectra, $a_i$ represents the individual elements of $\mathbf{a}$, $\mathbf{E}$ is a vector of size \emph{W}x1 representing additive noise and $\epsilon$ is the additive noise term for each spectrum \cite{manolakis2001hyperspectral}.

\begin{equation}
\label{eq:LMM_equation}
    \mathbf{y} = \sum_{i=1}^{N}a_i\mathbf{s_i} + \epsilon = \mathbf{Sa} + \mathbf{E}
\end{equation}

Linear least squares regression assumes the linear relationship in Equation \ref{eq:LMM_equation} and estimates the abundances $\mathbf{a}$ by finding abundances that minimize the sum of squares error (equivalently, minimizing the root mean squared error).  The formula for these estimated abundances is provided in Equation \ref{eq:LMM_solution}.

\begin{equation}
\label{eq:LMM_solution}
    \mathbf{\hat{a}} = (\mathbf{S^TS})^{-1}\mathbf{S^Ty}
\end{equation}

The nonlinear mixture model is often more appropriate for microscopic or intimate mixture scenarios because it relaxes the linear assumption to account for the interactions among light materials in the scene \cite{rasti2023image}. This modeling approach assumes that light photons reflect off more than one material before returning to the sensor \cite{keshava2002spectral,dobigeon2013nonlinear}. The alunite-kaolinite intimate mixture in Figure \ref{fig:complexmix} can be modeled in this way. There are various types of nonlinear mixture models, including bilinear models, the Hapke model and neural networks. Selection of approach depends on the composition of the pixel and the goal for unmixing. Though nonlinear models have the potential to achieve higher accuracy when this phenomena is present, certain nonlinear models demand information that is not easily obtainable by researchers \cite{combe2008analysis}.

\begin{figure}[h!]
    \centering
    \includegraphics[width=\columnwidth]{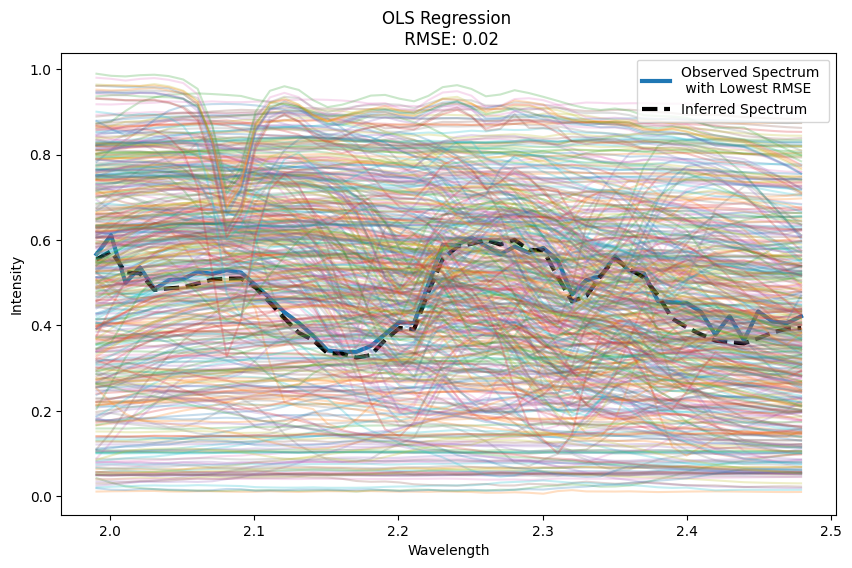}
    \caption{Depiction of unmixing results from OLS. The bold blue spectrum represents the observed pixel. The dashed spectrum represents the inferred model. The non-bold colored spectra represent the material included in the inferred model.}
    \label{fig:OLS challenge} 
\end{figure}

\subsection{Challenge in Unmixing}

The spectrum in an image for a pure material varies due to a number of factors. Factors intrinsic to the material include variation in particle size, surface roughness, variation over time from oxidation or UV-exposure.  For man-made materials, variation can occur from variation in manufacturing processes and ingredients.  For vegetation, spectra will vary with factors including health, age, season, soil conditions. Factors creating variation due to collection include variation in lighting intensity as well as angle and  atmospheric effects (when not fully corrected for)  \cite{borsoi2021spectral,combe2008analysis}. Often what is considered a `pure material' is really a class of materials with specific varying chemistry. For example, Ancylite is a group of hydrous strontium carbonate minerals which can contain varying amounts of cerium, lanthanum and minor amounts of other rare-earth elements (REEs). The common USGS specLib07b spectral library contains a single spectrum for Ancylite (Ancylite\_REE WS321 Darby MT).  It is not clear what the impact of variation in REE present is on the spectrum. In addition, naturally occurring minerals often have impurities and variations such as flocculation or layering of crystals that impact the resulting spectra.  The challenge of spectral variability creates a need for extensive spectral libraries incorporating the range of variation for priority materials to improve identification \cite{borsoi2020data}. Solving for the coefficients in a least squares regression problem with a large spectral library --- when there are more spectra in the library than bands --- results in an ill-posed inference problem \cite{borsoi2020data} requiring the inversion of a non-invertible matrix in Equation \ref{eq:LMM_equation}.

Ordinary least squares (OLS) regression often fails to accurately detect endmembers when materials are intimately mixed. Figure \ref{fig:OLS challenge} shows the result of unmixing a pixel from an image over the Cuprite Hills area using an OLS Python sklearn.linear\_model package, LinearRegression, with the USGS specLib07b library. The inferred pixel spectrum includes spectra from the entire spectral library -- represented by the non-bold colored spectra -- to model the observed pixel spectrum. Modeling accuracy is displayed through the fit of the plotted observed and inferred spectra and assessed using root mean squared error (RMSE). In the plot, the observed and inferred spectra have a very close fit. However, these results are erroneous and impractical because the entire spectral library of materials is not present in the pixel nor can there exist negative abundances values (which was needed to achieve this close model fit) for materials present in the pixel. The problem of a noninvertable matrix was handled computationally within the software. 

The problem demonstrated in Figure \ref{fig:OLS challenge} involves over-fitting by incorporating many irrelevant library spectra in the model.  This problem can be reduced or mitigated by incorporating constraints, regularization, and Bayesian methods.

A basic practical constraint that should almost always be imposed on the the linear mixture model is nonnegative abundance values. The implementation of the nonnegative constraint with least squares regression is frequently referred to as nonnegative least squares (NNLS) regression. Sometimes, the abundances are required to sum to one (interpreting them as physical percentage of the pixel area) as a constraint. Because of the variation in spectra, this constraint is often detrimental to results.

Sparse unmixing -- a common regularization approach -- involves the identification of the pure materials in a pixel using a spectral library by assuming the number of materials present in the pixel is small, often a substantially smaller subset of the materials in the library \cite{bioucas2010alternating}. Typically, sparsity is enforced through the use of a penalty on the model coefficients in an effort to shrink the abundances of unlikely substances. LASSO regression is an extension of OLS with an $L_1$ penalty norm on the coefficients (shown in Equation \ref{eq:LASSO_solution}) \cite{bedoui2020bayesian,iordache2013collaborative}. Ridge regression --- also an extension of OLS \cite{bedoui2020bayesian,iordache2013collaborative} --- enforces the $L_2$ penalty norm, but it is not considered a sparse technique as it encourages small coefficients rather than coefficients that are equal to zero. Researchers have achieved unmixing success by using LASSO regression and various combinations of LASSO with other techniques, because LASSO regression shrinks abundances to zero -- thereby ensuring sparsity. Moreover, the combination of $L_1$ and $L_2$ penalty norm on the coefficients, known as ElasticNet (shown in Equation \ref{eq:ElasticNet_solution}), has been useful in sparse unmixing.

\begin{equation}
\label{eq:LASSO_solution}
    \text{ arg min}_{a} \| y - \mathbf{Sa} \|_2^2 + \lambda \| \mathbf{a} \|_1
\end{equation}

\begin{equation}
\label{eq:ElasticNet_solution}
    \text{arg min}_{\mathbf{a}} \| y - \mathbf{Sa} \|_2^2 + \lambda_1 \| \mathbf{a} \|_1 + \lambda_2 \| \mathbf{a} \|_2^2
\end{equation}

Unsurprisingly, there has been successes in unmixing nonlinear mixtures using Bayesian regression as well \cite{altmann2015bayesian,dobigeon2008bayesian}. While sparse regression directly imposes a sparsity constraint on the coefficients, Bayesian regression uses of prior distributions on the model that encourage coefficients to be zero or near-zero. For instance, the $L_1$ penalty norm in LASSO regression is the equivalent of implementing a Laplace prior on the model abundances using Bayesian regression \cite{bedoui2020bayesian}. Thus, many sparse methods can be derived as Bayesian methods with appropriate choice of priors and assumptions.

Linear models can and have been implemented to unmix intimate mixtures \cite{schmidt2014minerals,themelis2011novel}. While the RMSE is not as low as can be achieved with nonlinear methods, the inferred constituent spectra in the linear model are often sufficient in accuracy for material identification. Simple nonlinear models can be created from the linear methods by adding quadratic (or higher order) terms for the potential constituent spectra. This increases the excess-variance problem and thus the need for regularization and sparsity.  

\subsection{Contribution}

There is limited research comparing the unmixing performance of sparse regression, feature search strategies, Bayesian regression, and nonlinear models for intimate mixtures. In fact, there is not a standard paradigm for evaluating performance of unmixing comparable to ROC curves for target detection.  Is unmixing successful only if it determines the right combination of constituents, and if so how do we evaluate unmixing on an image without knowing the correct combination at every pixel?  How much should determining the correct abundances matter?  If  unmixing determines a material that is chemically similar to, but not exactly, the believed-true constituent, is this better than determining an irrelevant material is present? Is a model wrong if it selects a correlated confuser rather than the actual material present in a model, and how much should this matter if the correlated confuser is chemically similar to the known true constituent?

Certainly, some of these questions are situation-dependent.  If the goal is to identify locations containing any variant of Ancylite then confusing Ancylite-La and Ancylite-Ce is not a problem.  In this paper we provide a framework for evaluating these for the problem of unmixing minerals. We also explore the degree to which the various unmixing methods perform relative to these considerations -- for example, which methods are likely to select multiple members of a material class versus selecting a single representative.

In this study, we compare the performance of LASSO regression, ElasticNet, Depth-First Search Feature Selection, BMA, and Nonlinear Quadratic Modeling. Because the base model of the feature search strategy, and BMA can be modified, the results will deviate between approach. We aim to identify the highest performing unmixing approach by testing these sparse techniques against their feature selection and ensambled version. The goal of this study is not to merely benchmark but also provide a discussion of the phenomena conflating spectral variability. We will provide a taxonomy of the physical-chemical attributes that aid or thwart positive detection of the target minerals.

The following is an overview of the succeeding sections of this paper. Section \ref{related work} details the research which motivated the methods pursued in this study. Following this literature review, in Section \ref{methods} we explain the unmixing techniques employed within our methodology. Section \ref{results} describes the results we obtained using these techniques. Finally, in Sections \ref{discussion} and \ref{conclusion}, we discuss the implications and conclusions derived from our results as well as ideas for further research.

\section{RELATED WORK}
\label{related work}
Combe et al. authored a paper focused on linearly unmixing two separate intimate mixture scenarios. One of the datasets was derived within a laboratory environment of synthesized Martian minerals. The other dataset was comprised of minerals collected from the well studied Cuprite Hills mining site in Nevada. With both datasets, the authors were able to detect the main materials contributing to the scene as well as the spectral characteristics of the minerals leading to positive detection \cite{combe2008analysis}.

Itoh and Parente coauthored a study exploring sparse regression in nonlinear mixtures. Their research compared the performance of a sparse technique (LASSO regression) with other unmixing methods, including NNLS. They found that applying an abundance threshold—set at 0.1 in their study—substantially improved the detection of pure materials, to the point where NNLS could outperform LASSO in endmember determination \cite{itoh2016performance}. This result is logical, as the threshold acts as a form of regularization, filtering out low-abundance components and thereby refining the detection of pure materials and improving overall performance.

In both Li et al. and Iordache et al., the authors discussed methodologies for sparse unmixing using penalty norms \cite{li2019local}. Li et al. proposed a new algorithm incorporating the \(L_{2,1}\) norm to preserve spectral information for pixels in the same region. Iordache et al. provided a comparison paper of an algorithm incorporating the \(L_{2,1}\) norm, CLSUnSAL, proposed in a preceding paper by Bioucas-Dias and Figueiredo, and other known sparse regression algorithms \cite{bioucas2011alternating,iordache2013collaborative}.

Dash and Liu's paper discussed variations in feature selection use cases, defining feature selection as a ``search procedure" with iterative properties for identifying the most representative characteristics of the observed data \cite{dash1997feature}. Gault et al. and Winter et al. both implemented iterative unmixing approaches in their studies. Winter et al. emphasize that iterative approaches are useful for adaptively reducing error for model selection. Gault et al. compare the unmixing performance of iterative techniques --- LASSO regression, stepwise regression, and fully constrained least squares --- with non-iterative methods like unconstrained least squares \cite{gault2016comparing,winter2003examining}.

Rankin et al. proposed an approach aimed at long wave infrared image analysis for target detection. In their work, they presented a methodology incorporating the fully constrained least squares model and BMA for material abundance and temperature estimation. The authors' approach incorporated equal priors and evaluated model likelihoods using bayesian information criteria (BIC) scores \cite{rankin2017spectral}.

In another paper aimed at model selection, M\"uller et al. discussed bayesian methods and information criteria approaches. Unlike Rankin et. al, M\"uller et al. did not implement BMA. They compared Akaike Information criteria (AIC), BIC and other bayesian approaches to regularization regression methods \cite{muller2013model}.

Themilis et al. presented a hierarchical bayesian approach to perform sparse unmixing with nonnegativity constraints. Specifically, their model assigned a two-level hierarchical prior on parameters encouraging sparsity in the abundance estimates. The authors also demonstrated their model was equivalent to LASSO regression \cite{themelis2011novel}. 

\begin{table*}[h!]
\centering
\resizebox{\textwidth}{!}{%
    \begin{tabular}{lcccccccccc}
    \hline
    & \makecell{LASSO} & \makecell{ElasticNet} & \makecell{DFS\\LASSO} & \makecell{DFS\\ElasticNet} & \makecell{BMA\\NNLS} & \makecell{BMA\\LASSO} & \makecell{BMA\\ElasticNet} & \makecell{BMA-Q\\NNLS} & \makecell{BMA-Q\\LASSO} & \makecell{BMA-Q\\ElasticNet} \\
    \hline
    Alpha & 0.0004 & 0.001 & 0.0001 & 0.001 & - & 0.0001 & 0.0001 & - & 0.0001 & 0.001 \\
    \hline
    \end{tabular}
}
\caption{Presentation of alpha parameters for each unmixing technique. The parameter $\alpha$ controls the regularization strength for techniques that use regularization, while `-' indicates methods where $\alpha$ is not applicable.}
\label{parametertable}
\end{table*}

\section{METHODS}
\label{methods}

\begin{figure*}[hp] 
\begin{center}
\begin{subfigure}{.47\textwidth}
\includegraphics[width=\textwidth]{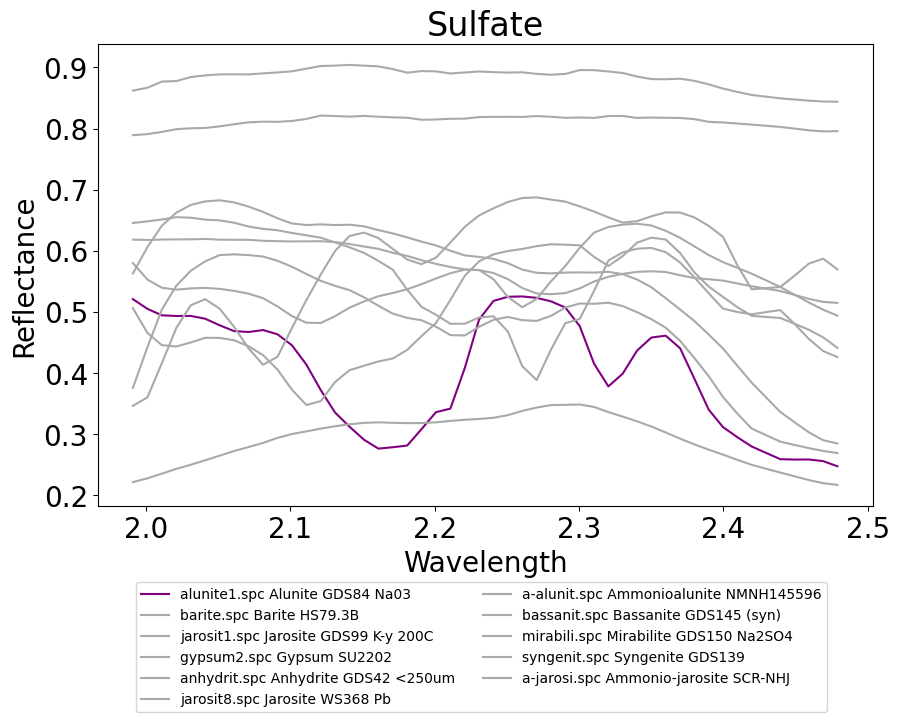}
\caption{Chemical Formula: $\text{SO}_4$}
\label{fig:Sulfate}
\end{subfigure}
\begin{subfigure}{.47\textwidth}
\includegraphics[width=\textwidth]{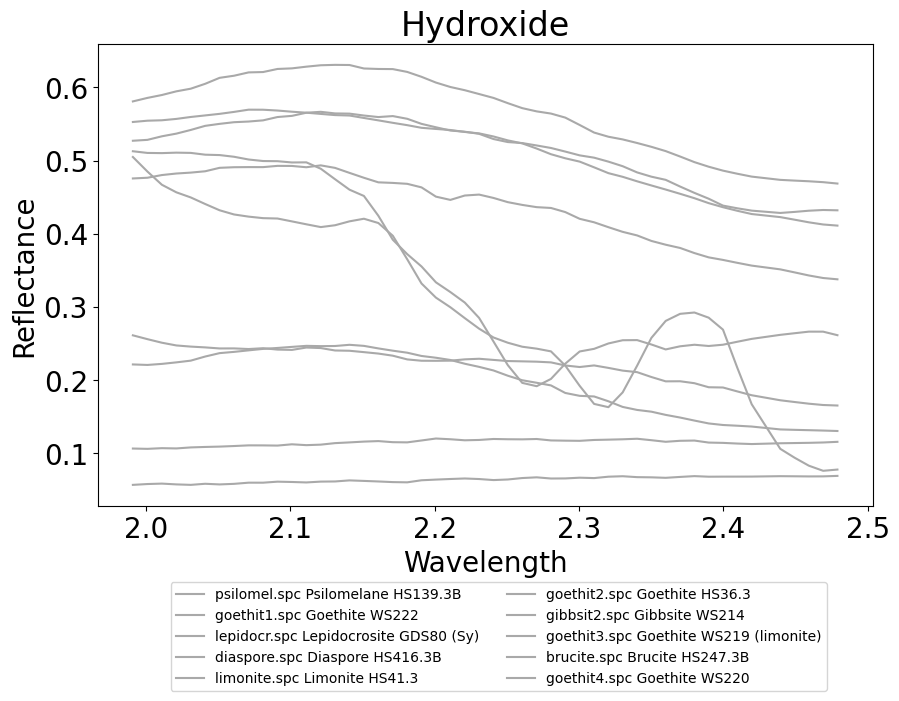}
\caption{Chemical Formula: OH$-$}
\label{fig:Hydroxide}
\end{subfigure}
\begin{subfigure}{.47\textwidth}
\includegraphics[width=\textwidth]{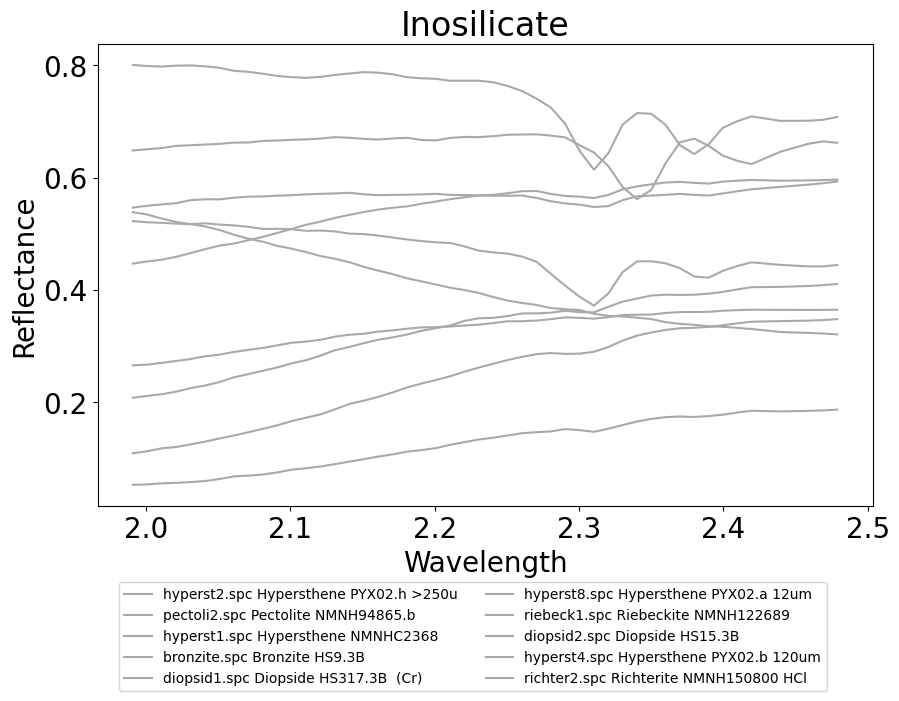}
\caption{Chemical Formula: $\text{SiO}_3$}
\label{fig:Inosilicate}
\end{subfigure}
\begin{subfigure}{.47\textwidth}
\includegraphics[width=\textwidth]{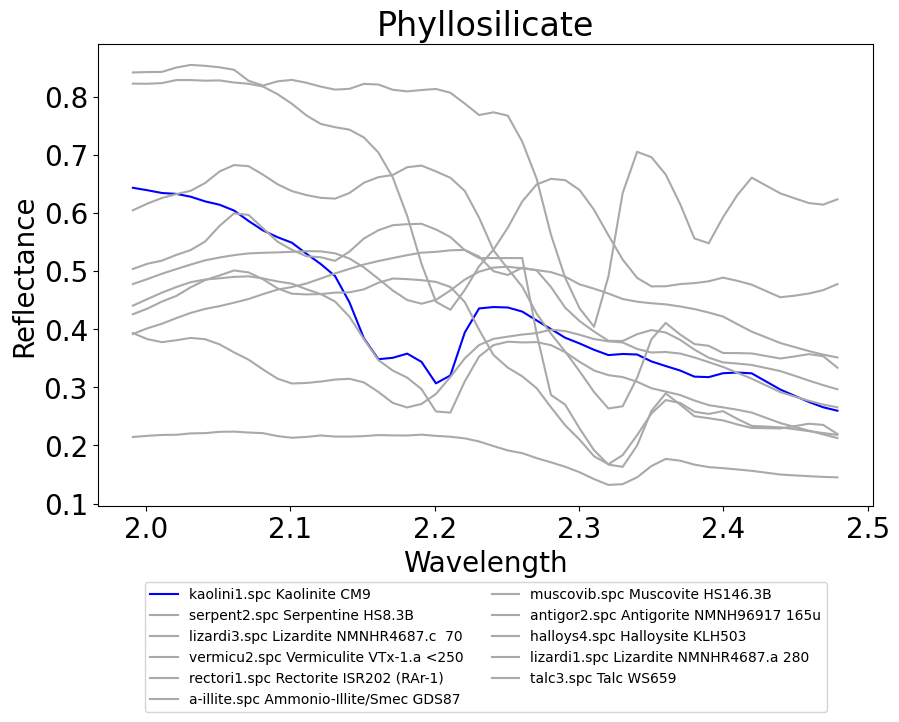}
\caption{Chemical Formula: $\text{Si}_2\text{O}_5$}
\label{fig:Phyllosilicate}
\end{subfigure}
\begin{subfigure}{.47\textwidth}
\includegraphics[width=\textwidth]{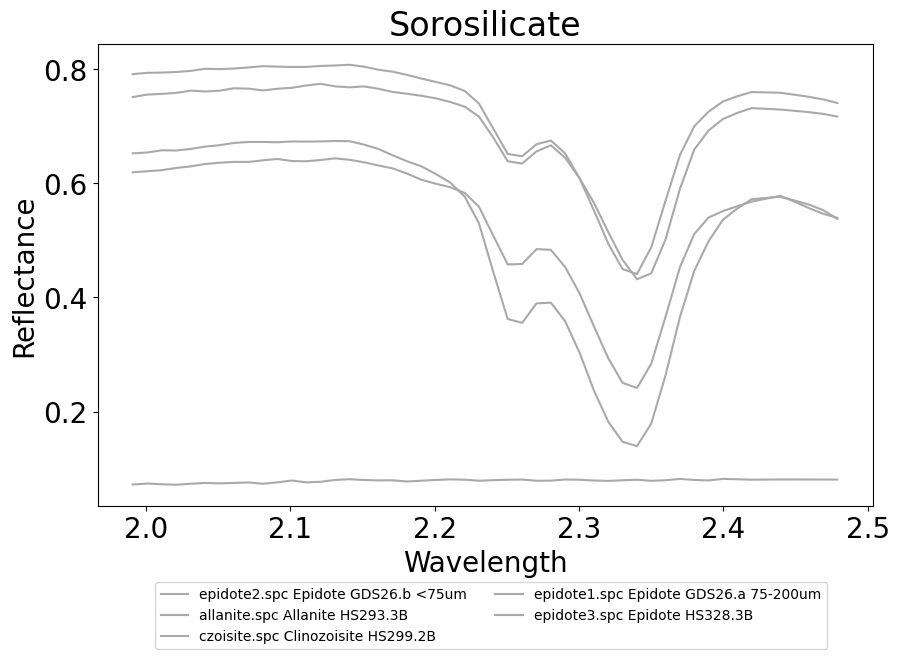}
\caption{Chemical Formula: $\text{Si}_2\text{O}_7$}
\label{fig:Sorosilicate}
\end{subfigure}
 \hspace{.5cm}
\begin{subfigure}{.47\textwidth}
\includegraphics[width=\textwidth]{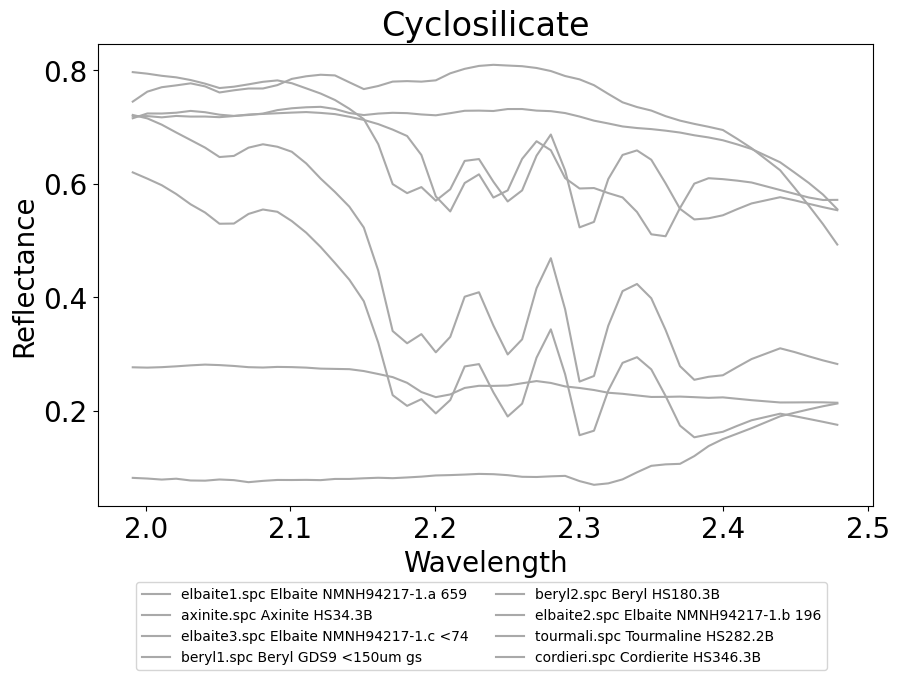}
\caption{Chemical Formula: $\text{SiO}_3$}
\label{fig:Cyclosilicate}
\end{subfigure}
\end{center}

\caption[example] { \label{fig:physical-chemical 1} 
Display of mineral types sulfate, hydroxide, and inosilicate, phyllosilicate, sorosilicate, and cyclosilicate. Figure \ref{fig:Sulfate} shows the target mineral, alunite, highlighted in purple. Figure \ref{fig:Phyllosilicate} shows the target mineral, kaolinite highlighted in blue.}
\end{figure*}

\subsection{Dataset}
Our observed data was comprised of hyperspectral image pixels captured by the Airborne Visible/Infrared Imaging Spectrometer (AVIRIS) over Cuprite Hills, Nevada. Using the Python software package ``Hyperspectralpy", we selected pixels from the image to form our regions of interest (ROIs) \cite{Basener_2022b}. Two ROIs were selected: an alunite hill and kaolinte region.

The spectral library was obtained from the United States Geological Survey (USGS). This library is comprised of 481 spectral signatures, mapped over 50 spectral bands, ranging from 2 to 2.5 $\mu$m. Many of the spectra within the library are variants within the same mineral class or are variations of the same mineral (a mineral has a fixed chemical formula, but the spectra can vary as described previously). 

\begin{table*}[h!]
\centering

\vspace{2em}

\begin{subtable}{\textwidth}
\centering
\resizebox{\textwidth}{!}{%
\begin{tabular}{llccccc}
\hline
\textbf{Category} & \textbf{Technique} & \textbf{RMSE Mean} & \textbf{RMSE Std} & \textbf{Model Size} & \textbf{Runtime (s)} & \textbf{Detection} \\
\hline
\multirow{2}{*}{Regularization Approaches} & LASSO & 0.1310 & 0.0892 & 4.3740 & 0.0050 & 0.9593 \\
                                   & ElasticNet & 0.1389 & 0.0870 & 7.0650 & 0.0033 & 0.9593 \\
\hline
\multirow{2}{*}{Iterative Approaches} & DFS LASSO & 0.0868 & 0.0737 & 2.0894 & 1.0946 & 0.9268 \\
                                       & DFS ElasticNet & 0.0857 & 0.0726 & 2.1870 & 1.0772 & 0.9268 \\
\hline
\multirow{3}{*}{Ensembling Approaches} & BMA NNLS & 0.0519 & 0.0281 & 2.7724 & 0.0487 & 0.8862 \\
                                       & BMA LASSO & 0.0538 & 0.0206 & 2.0732 & 0.6012 & 0.7398 \\
                                       & BMA ElasticNet & 0.0525 & 0.0183 & 2.1870 & 0.6168 & 0.7480 \\
\hline
\multirow{3}{*}{Nonlinear Approaches} & BMAQ NNLS & 0.2743 & 0.1438 & 3.1382 & 0.0760 & 0.8455 \\
                                      & BMAQ LASSO & 0.2452 & 0.1919 & 2.4309 & 0.6928 & 0.5203 \\
                                      & BMAQ ElasticNet & 0.2696 & 0.1273 & 2.3008 & 0.6403 & 0.3333 \\
\hline
\end{tabular}%
}
\caption{Metrics for unmixing techniques for the alunite ROI.}
\label{subtable:unmixing_techniques_1}
\end{subtable}

\begin{subtable}{\textwidth}
\centering
\resizebox{\textwidth}{!}{%
\begin{tabular}{llccccc}
\hline
\textbf{Category} & \textbf{Technique} & \textbf{RMSE Mean} & \textbf{RMSE Std} & \textbf{Model Size} & \textbf{Runtime (s)} & \textbf{Detection} \\
\hline
\multirow{2}{*}{Regularization Approaches} & LASSO & 0.1348 & 0.0777 & 6.8917 & 0.0143 & 0.9667 \\
                                   & ElasticNet & 0.1384 & 0.0788 & 12.5167 & 0.0099 & 1.0000 \\
\hline
\multirow{2}{*}{Iterative Approaches} & DFS LASSO & 0.0732 & 0.0437 & 2.3167 & 1.3576 & 0.7583 \\
                                       & DFS ElasticNet & 0.0733 & 0.0426 & 2.4500 & 1.3540 & 0.7917 \\
\hline
\multirow{3}{*}{Ensembling Approaches} & BMA NNLS & 0.0630 & 0.0268 & 3.6750 & 0.0730 & 0.6667 \\
                                       & BMA LASSO & 0.0484 & 0.0108 & 2.7083 & 0.6422 & 0.6833 \\
                                       & BMA ElasticNet & 0.0493 & 0.0124 & 2.6917 & 0.6552 & 0.6833 \\
\hline
\multirow{3}{*}{Nonlinear Approaches} & BMAQ NNLS & 0.3863 & 0.1208 & 4.6250 & 0.1529 & 0.6167 \\
                                      & BMAQ LASSO & 0.4923 & 0.1832 & 2.3667 & 0.6293 & 0.3000 \\
                                      & BMAQ ElasticNet & 0.4125 & 0.0962 & 2.5417 & 0.6891 & 0.3000 \\
\hline
\end{tabular}%
}
\caption{Metrics for unmixing techniques for the kaolinite ROI.}
\label{subtable:unmixing_techniques_2}
\end{subtable}

\caption{Comparison of unmixing techniques with metrics including RMSE mean, RMSE standard deviation, model size, runtime, and detection rate for two datasets.}
\label{tab:unmixing_techniques_combined}

\end{table*}

\subsection{Unmixing Techniques}
LASSO regression and ElasticNet are frequently implemented regularization regression techniques. We incorporated a basic version of both techniques to identify their individual advantages as they both encourage sparse solutions. Python's Sklearn ``LASSO" and ``ElasticNet" packages were employed for the basic implementation.

For the iterative approaches, we independently developed (without the use of a Python package) two variations of DFS feature selection strategies. Our feature selection variations implemented LASSO regression and ElasticNet. We incorporated these variations to simultaneously enforce sparsity and reduce modeling error.

LASSO and ElasticNet were also incorporated in a basic BMA implementation and BMA with quadratic terms, BMA-Q. We compared these sparse versions of BMA and BMA-Q to non sparse versions using NNLS. In total, we generated six different comparisons using BMA: BMA NNLS, BMA LASSO, BMA ElasticNet, BMA-Q NNLS, BMA-Q LASSO, and BMA-Q ElasticNet.  The BMA-Q techniques included interactions of materials up to the 2nd order. We independently developed all of the BMA variations rather than using a Python package.

Table \ref{parametertable} presents the alpha parameters used for each method to achieve the unmixing results. Each technique is evaluated based on RMSE, model size, computation time, and detection percentage. The average modeling error is reported alongside its variability, measured by the standard deviation (std). Computation time refers to the average time required for each technique to unmix a single pixel within the ROI. Detection percentage indicates the proportion of inferred models in the ROI that successfully contain the target mineral. Unmixing performance may vary based on factors such as the target mineral, pixel location, and the composition of the mineral mixture. This comparison aims to determine the most effective techniques in certain scenarios. 

\subsection{Physical-chemical Taxonomy}
Every mineral is comprised of a chemical make up which dictates their physical appearance and texture. Mineralogists use this chemical structure to define and categorize these substances. Conveniently, these mineral types follow certain spectral patterns. We not only examined the mineral category of our target minerals, alunite and kaolinite, but we also identified the categories commonly detected with the target mineral type. Figure \ref{fig:physical-chemical 1} displays six mineral types: sulfate, halide, inosilicate, phyllosilicate, sorosilicate, and cyclosilicate. The spectral library encompassed approximately twenty categories, but the mineral types referenced here frequently occur in the unmixing results. Each of the subplots in Figure \ref{fig:physical-chemical 1} displays the pattern  of a few of the minerals within a certain category from the spectral library.

\section{RESULTS}
\label{results}

\begin{figure*}[hp!] 
\centering

\begin{subfigure}{\textwidth}
        \includegraphics[width=.9\textwidth]{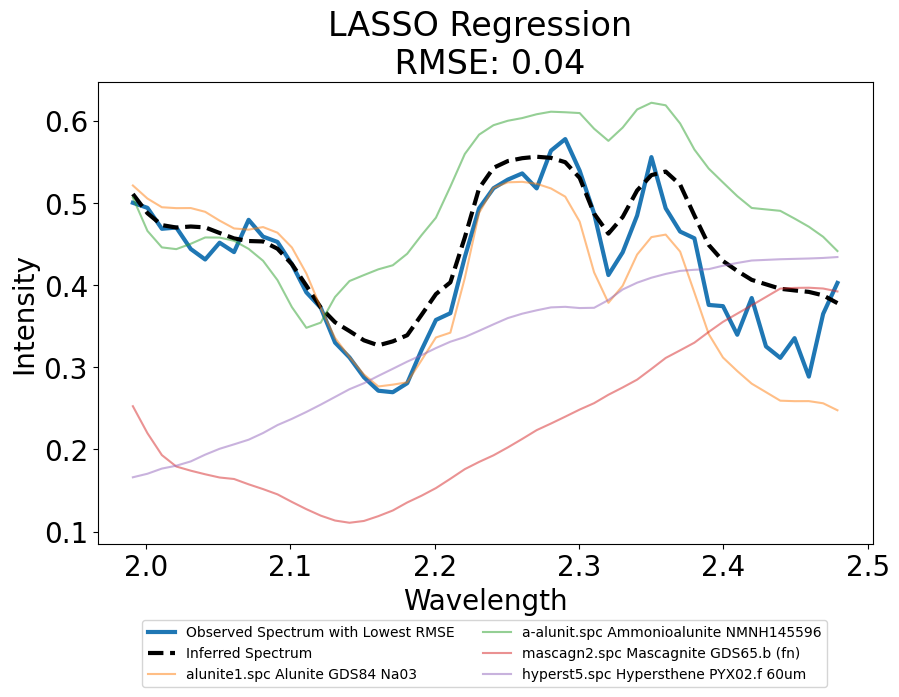}
\end{subfigure}\\

\begin{subfigure}{\textwidth}
        \centering 
        \resizebox{\textwidth}{!}{
        \begin{tabular}{lllc}
        \toprule
        \textbf{Name} & \textbf{Category} & \textbf{Formula} & \textbf{Abundance} \\ \midrule
        \rowcolor{yellow} \texttt{alunite1.spc Alunite GDS84 Na03} & Sulfate & $(\text{Na}, \text{K})\text{Al}_3(\text{SO}_4)_2(\text{OH})_6$ & 0.6641 \\
        \rowcolor{yellow} \texttt{a-alunit.spc Ammonioalunite NMNH145596} & Sulfate & $(\text{NH}_4)\text{Al}_3(\text{SO}_4)_2(\text{OH})_6$ & 0.1950 \\
        \texttt{mascagn2.spc Mascagnite GDS65.b (fn)} & Sulfate & $(\text{NH}_4)_2\text{SO}_4$ & 0.1667 \\
        \texttt{hyperst5.spc Hypersthene PYX02.f 60um} & Inosilicate & $(\text{Mg},\text{Fe}^{2+})_2\text{Si}_2\text{O}_6$ & 0.1430 \\ 
        \bottomrule
        \end{tabular}
        }
    \end{subfigure}
    \caption{\label{fig:LassoAlunite} Alunite Table and Plot. Displays the inferred spectrum using LASSO regression with the most common minerals and their average abundances. Rows containing alunite are highlighted in yellow.}
\end{figure*}

\begin{figure*}[hp!]
\centering
\begin{subfigure}{.9\textwidth}
\includegraphics[width=\textwidth]{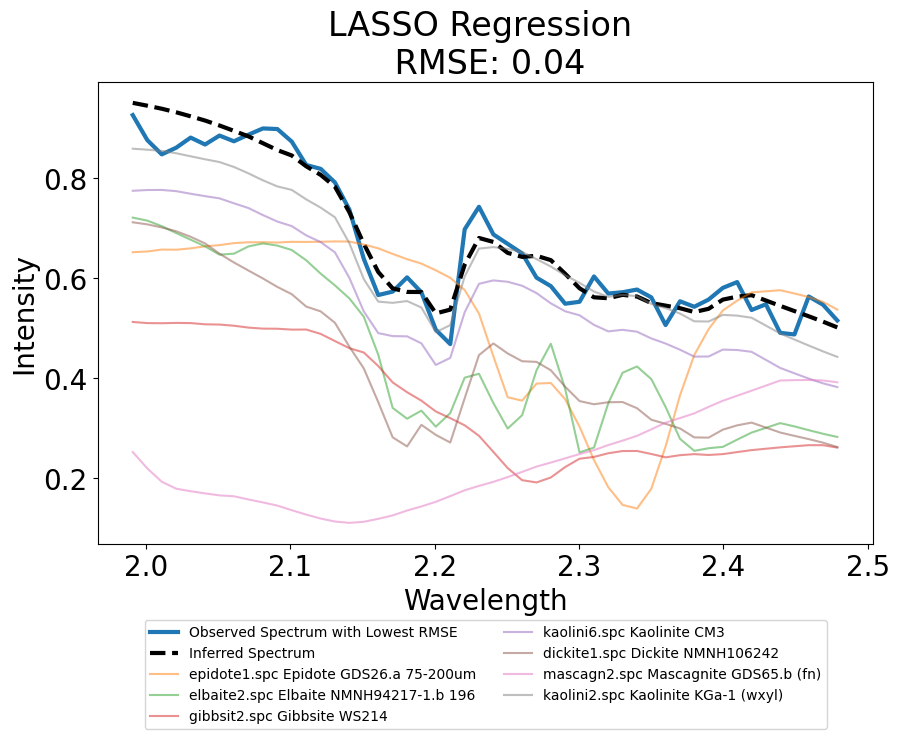}
\end{subfigure}

\begin{subfigure}{\textwidth}
        \centering
        \resizebox{\textwidth}{!}{
        \begin{tabular}{lllc}
        \toprule
        \textbf{Name} & \textbf{Category} & \textbf{Formula} & \textbf{Abundance} \\ \midrule
        \texttt{epidote1.spc Epidote GDS26.a} & Sorosilicate & $\text{Ca}_2(\text{Al,Fe}^{+3})_3(\text{SiO}_4)_3(\text{OH})$ & 0.0932 \\
        \texttt{elbaite2.spc Elbaite NMNH94217-1.b} & Cyclosilicate & $\text{Na}(\text{Li,Al})_3\text{Al}_6(\text{BO}_3)_3\text{Si}_6\text{O}_{18}(\text{OH})_4$ & 0.0966 \\
        \texttt{gibbsit2.spc Gibbsite WS214} & Hydroxide & $\text{Al}(\text{OH})_3$ & 0.1172 \\
        \rowcolor{cyan} \texttt{kaolini6.spc Kaolinite CM3} & Phyllosilicate & $\text{Al}_2\text{Si}_2\text{O}_5(\text{OH})_4$ & 0.3362 \\
        \texttt{dickite1.spc Dickite NMNH106242} & Phyllosilicate & $\text{Al}_2\text{Si}_2\text{O}_5(\text{OH})_4$ & 0.2205 \\
        \texttt{mascagn2.spc Mascagnite GDS65.b} & Sulfate & $(\text{NH}_4)_2\text{SO}_4$ & 0.1171 \\
        \rowcolor{cyan} \texttt{kaolini2.spc Kaolinite KGa-1 (wxyl)} & Phyllosilicate & $\text{Al}_2\text{Si}_2\text{O}_5(\text{OH})_4$ & 0.3645 \\ 
        \bottomrule
        \end{tabular}
        }
    \end{subfigure}
\caption{ \label{fig:LassoKaolinite} 
Kaolinite Table and Plot. Displays the inferred spectrum using LASSO regression with the most common minerals and their average abundances. Rows containing kaolinite are highlighted in cyan.}
\end{figure*}

\begin{figure*}[hp!]
\centering
\begin{subfigure}{\textwidth}
\includegraphics[width=.9\textwidth]{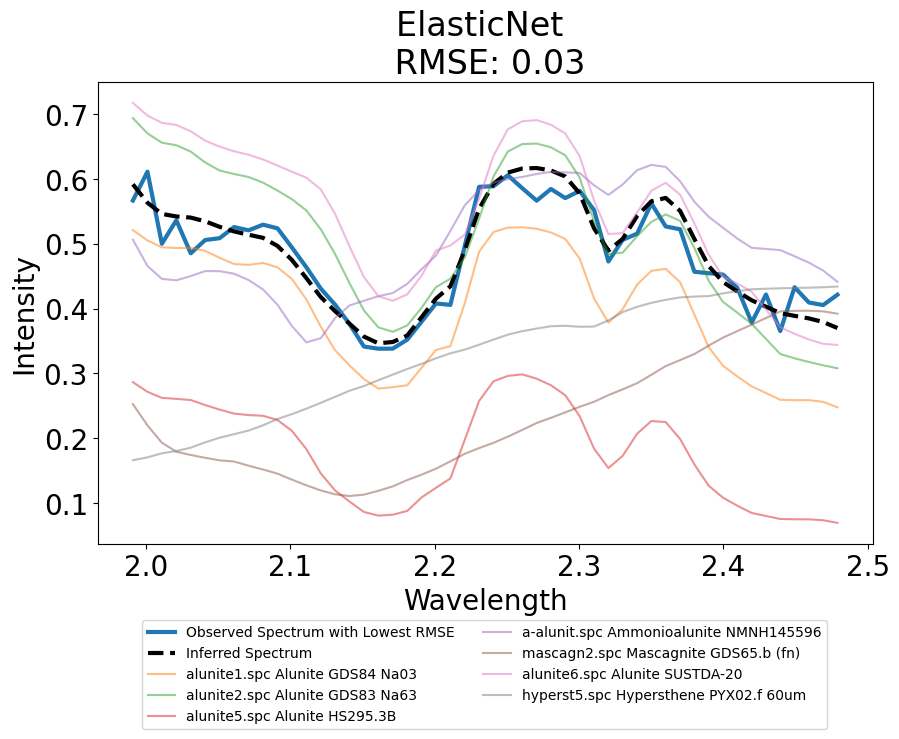}
\end{subfigure}

\begin{subfigure}{\textwidth}
        \centering
        \resizebox{\textwidth}{!}{
        \begin{tabular}{lllc}
        \toprule
        \textbf{Name} & \textbf{Category} & \textbf{Formula} & \textbf{Abundance} \\ \midrule
        \rowcolor{yellow} \texttt{alunite1.spc Alunite GDS84 Na03} & Sulfate & $(\text{Na}, \text{K})\text{Al}_3(\text{SO}_4)_2(\text{OH})_6$ & 0.3596 \\
        \rowcolor{yellow} \texttt{alunite2.spc Alunite GDS83 Na63} & Sulfate & $(\text{Na}, \text{K})\text{Al}_3(\text{SO}_4)_2(\text{OH})_6$ & 0.1706 \\
        \rowcolor{yellow} \texttt{alunite5.spc Alunite HS295.3B} & Sulfate & $(\text{Na}, \text{K})\text{Al}_3(\text{SO}_4)_2(\text{OH})_6$ & 0.2280 \\
        \rowcolor{yellow} \texttt{a-alunit.spc Ammonioalunite NMNH145596} & Sulfate & $(\text{NH}_4)\text{Al}_3(\text{SO}_4)_2(\text{OH})_6$ & 0.2285 \\
        \texttt{mascagn2.spc Mascagnite GDS65.b (fn)} & Sulfate & $(\text{NH}_4)_2\text{SO}_4$ & 0.1429 \\
        \rowcolor{yellow} \texttt{alunite6.spc Alunite SUSTDA-20} & Sulfate & $(\text{Na}, \text{K})\text{Al}_3(\text{SO}_4)_2(\text{OH})_6$ & 0.0811 \\
        \texttt{hyperst5.spc Hypersthene PYX02.f 60um} & Inosilicate & $(\text{Mg}, \text{Fe}^{+2})_2\text{Si}_2\text{O}_6$ & 0.0640 \\ 
        \bottomrule
        \end{tabular}
        }
    \end{subfigure}
\caption{ \label{fig:ElasticNetAlunite} 
Alunite Table and Plot. Displays the inferred spectrum using ElasticNet with the most common minerals and their average abundances. Rows containing alunite are highlighted in yellow.}
\end{figure*}

\begin{figure*}[hp!]
\centering
\begin{subfigure}{\textwidth}
\includegraphics[width=.9\textwidth]{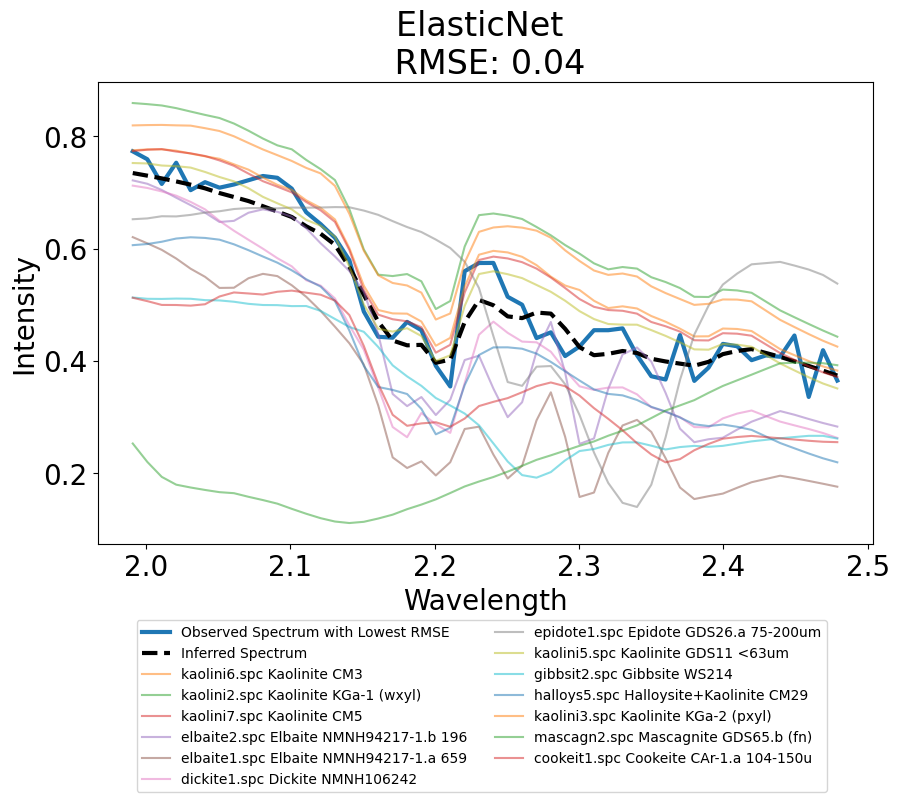}
\end{subfigure}

\begin{subfigure}{\textwidth}
        \centering
        \resizebox{\textwidth}{!}{
        \begin{tabular}{lllc}
        \toprule
        \textbf{Name} & \textbf{Category} & \textbf{Formula} & \textbf{Abundance} \\ \midrule
        \rowcolor{cyan} \texttt{kaolini6.spc Kaolinite CM3} & Phyllosilicate & $\text{Al}_2\text{Si}_2\text{O}_5(\text{OH})_4$ & 0.1087 \\
        \rowcolor{cyan} \texttt{kaolini2.spc Kaolinite KGa-1 (wxyl)} & Phyllosilicate & $\text{Al}_2\text{Si}_2\text{O}_5(\text{OH})_4$ & 0.1046 \\
        \rowcolor{cyan} \texttt{kaolini7.spc Kaolinite CM5} & Phyllosilicate & $\text{Al}_2\text{Si}_2\text{O}_5(\text{OH})_4$ & 0.1041 \\
        \texttt{elbaite2.spc Elbaite NMNH94217-1.b 196} & Cyclosilicate & $\text{Na}(\text{Li,Al})_3\text{Al}_6(\text{BO}_3)_3\text{Si}_6\text{O}_{18}(\text{OH})_4$ & 0.0759 \\
        \texttt{elbaite1.spc Elbaite NMNH94217-1.a 659} & Cyclosilicate & $\text{Na}(\text{Li,Al})_3\text{Al}_6(\text{BO}_3)_3\text{Si}_6\text{O}_{18}(\text{OH})_4$ & 0.0597 \\
        \texttt{dickite1.spc Dickite NMNH106242} & Phyllosilicate & $\text{Al}_2\text{Si}_2\text{O}_5(\text{OH})_4$ & 0.1696 \\
        \texttt{epidote1.spc Epidote GDS26.a 75-200um} & Sorosilicate & $\text{Ca}_2(\text{Al,Fe}^{+3})_3(\text{SiO}_4)_3(\text{OH})$ & 0.0951 \\
        \rowcolor{cyan} \texttt{kaolini5.spc Kaolinite GDS11 <63um} & Phyllosilicate & $\text{Al}_2\text{Si}_2\text{O}_5(\text{OH})_4$ & 0.0576 \\
        \texttt{gibbsit2.spc Gibbsite WS214} & Hydroxide & $\text{Al}(\text{OH})_3$ & 0.0673 \\
        \rowcolor{cyan} \texttt{halloys5.spc Halloysite+Kaolinite CM29} & Phyllosilicate & Mixture of halloysite and kaolinite & 0.0738 \\
        \rowcolor{cyan} \texttt{kaolini3.spc Kaolinite KGa-2 (pxyl)} & Phyllosilicate & $\text{Al}_2\text{Si}_2\text{O}_5(\text{OH})_4$ & 0.0477 \\
        \texttt{mascagn2.spc Mascagnite GDS65.b (fn)} & Sulfate & $(\text{NH}_4)_2\text{SO}_4$ & 0.0885 \\
        \texttt{cookeit1.spc Cookeite CAr-1.a 104-150u} & Phyllosilicate & $\text{LiAl}_4(\text{Si}_3\text{Al})\text{O}_{10}(\text{OH})_8$ & 0.0416 \\ 
        \bottomrule
        \end{tabular}
        }
    \end{subfigure}
\caption{ \label{fig:ElasticNetKaolinite} 
Kaolinite Table and Plot. Displays the inferred spectrum using ElasticNet with the most common minerals and their average abundances. Rows containing kaolinite are highlighted in yellow.}
\end{figure*}

\subsection{Unmixing Performance}

Table \ref{tab:unmixing_techniques_combined} compares the unmixing performance of the LASSO regression, ElasticNet, DFS using LASSO, DFS using ElasticNet, BMA using NNLS, BMA using LASSO, BMA using ElasticNet, BMA-Q using NNLS, BMA-Q using LASSO, and BMA-Q using ElasticNet. The table shows the detection percentage, average model size, average computation time, average RMSE and the standard deviation of the RMSE. LASSO regression and ElasticNet performed better than all the other techniques in terms of computation time, and detection percentage. Both techniques had the largest models as well as the highest modeling error of the linear approaches. ElasticNet achieved slightly better accuracy when unmixing pixels for the kaolinite ROI. Overall, the two sparse approaches had similar results ranging from 95-100\% detection. 

The iterative approaches had fairly high accuracy in detecting the target minerals and also achieved low error rates with low variability for both ROIs. DFS feature selection achieved a higher detection percentage with the alunite ROI (above 90\% detection of the target mineral) than with the kaolinite ROI (detection rate between 75--80\%). Despite achieving the second highest detection rates, the computation time for DFS was the highest across all methods implemented. 

The detection percentage for BMA was lower compared to the sparse and iterative approaches; however, it achieved the lowest average error with the narrowest error spread among all techniques.  BMA NNLS achieved a better detection percentage than BMA LASSO and BMA ElasticNet for the alunite ROIs. Furthermore, detection percentages for the BMA techniques were substantially lower with the kaolinite ROI than the alunite ROI. Also, it is interesting to note that the computation time for BMA NNLS was much faster than that for BMA LASSO and BMA ElasticNet. Even with the lower detection rates, the computed error for these linear ensambling techniques was lower than the sparse and iterative approaches. This means there probably exists an inferred model from both BMA LASSO and BMA ElasticNet that resembles a close fit to at least one of the pixels.

The nonlinear approaches achieved the lowest performance in terms of detection accuracy. Also, the detection percentages were spread over wider ranges --- 33--84\% with the alunite ROI and 30-62\% with the kaolinite ROI. Similar to BMA NNLS, BMA-Q NNLS performed better than the LASSO and ElasticNet versions in terms of computation time, error and detection. The average unmixing error was lower with the alunite ROI (RMSE Mean = 0.27, 0.25, 0.27) than with the kaolinite ROI (RMSE Mean = 0.40, 0.49, 0.41) but the variability of the BMA-Q modeling error was the highest across all the techniques (RMSE Std $>$ 0.09).

\subsection{Feature Taxonomy}
Figures \ref{fig:LassoAlunite} and \ref{fig:LassoKaolinite} show the inferred models from LASSO regression and Figures \ref{fig:ElasticNetAlunite} and \ref{fig:ElasticNetKaolinite} show ElasticNet for both ROIs. The figures display an inferred model generated from the --- technique specific --- most common minerals using the average estimated abundance for that mineral. This inferred model is plotted with the closet fitting observed pixel spectrum and the constituent library spectra. For example, Figure \ref{fig:LassoAlunite} shows alunite, ammonioalunite, mascagnite and hypersthene as the most common minerals included in the inferred model using LASSO regression. Their average abundances using this technique are listed in the table and used to plot their inferred spectrum. The table beneath the plot shows not only each mineral's estimated fractional abundance, but also each mineral's classification category and associated chemical formula. 

\section{DISCUSSION}
\label{discussion} 
The results from this taxonomy indicates that many of the minerals included in the inferred models were from the same chemical category or categories with similar spectral features as the target mineral. Mascanite was the common alternative sulfate incorporated in the model regardless of technique for the alunite ROI. Hypersthene -- a inosilicate mineral -- was the other frequently incorporated mineral. 

Similarly, the models for the kaolinite ROI included minerals from the same phyllosilicate category or a mineral from a chemical category with similar features in specific bands. Dicktie was a common phyllosilicate incorporated. Three other common mineral categories --- for the kaolinite ROI --- were cyclosilicate, sorosilicate and hydroxide. Models incorporated variations of elbaite from the cyclosilicate category. Epidote and gibbsite (respectively) were the most frequently incorporated sorosilicate and hydroxide.

\section{Conclusion}
\label{conclusion}
In this study we compared the performance of 10 unmixing techniques. We measured performance based on the proportion of inferred spectra that included the target mineral, the computed RMSE, average model size and the unmixing computation time. The sparse approaches preformed the best because they quickly detected the target mineral for almost every pixel. LASSO and ElasticNet had mildly better performance depending on the ROI, therefore we recommend implementing both methods for unmixing intimate mixtures. 

A notable observation from the ElasticNet unmixing results was the detection of multiple variations of the target mineral. Each of the techniques were able to positively detect alunite and kaolinite, but ElasticNet had several variation of the target material from the spectral library. ElasticNet involves both the L1 and L2 norm. Modifying the penalty term in ElasticNet caused the model to be less sparse than LASSO regression but sparse enough to allow only close variations. This inclusion of near-variations may reflect ElasticNet's ability to capture subtle spatial nuances and underlying phenomenology

Intra-class variations were not the only variability observed from the taxonomy. We observed inter-class variability as well. Minerals from different chemical categories and their slight variations were often included. Certain mineral categories resembled the target mineral category within specific bands and therefore were included in the model to obtain a closer model fit. If the target was not included in the model, it was commonly replaced with a mineral from the same chemical category.


\bibliography{refs} 
\bibliographystyle{IEEEbib} 

\appendix 
\section{Referenced code}
All the referenced and supporting code will be stored within the corresponding GitHub repository (\url{https://github.com/bakerjf1993/Hyperspectral-Unmixing/tree/main/2024%20IASIM%20Spectrochimica%20Acta}).

\end{document}